# Source and Channel Optimal Rate LDPC Code Design for one Sender in BE-MAC with Source Correlation

H. Tavakoli, Assistant Professor
Department of Electrical
Engineering, Faculty of Engineering
University of Guilan, Rasht, Iran
htavakoli@guilan.ac.ir

*Abstract*—In this paper, we present an extension of the semi-definite programming formulation of the optimal rate code design in single link Binary Erasure Channel (BEC) proposed by the authors to the Binary Erasure Multiple Access Channel (BE-MAC) with two sources correlation. This new way can be easily extended to the multiple access senders. Simulation results show the efficiency and effectiveness of the new approach in practice.

## I. INTRODUCTION

After forgetting LDPC codes, introduced by Gallager in [1], Mackey's paper causes some attentions to them [2]. LDPC code decoders are belief-propagation or sum-product [1,3]. These processes are mathematically explained by a nonlinear inequality, the so called Density Evolution (DE), which was first formulated by Richardson and Urbanke [3-5].

One can solve optimal rate problem to maximize the rate of the code subject to DE constraint. This problem is a semi-infinite optimization problem. It is worth mentioning that the DE constraint in BEC and BSC is a polynomial function [5-10] while in AWGN channel it is a nonlinear function.

We introduced an approach based on semidefinite programming formulation of the optimal rate problem in BEC which takes full responsibility to the all DE constraint instead of other methods [11-13].

In this paper, we extend the optimal rate problem from single user BEC to the multiple correlated user. To do so, we first introduce model, based on literature models. In addition to DE constraints, other considerations such as Slepian-Wolf [14], should be added to the model which makes it more complicated than the single user BEC mode. As the extension of the model to the more general case is straight. We restrict our model to a correlated user to another user. By solving the optimal rate problem for a given set parameter: eraser rate, check node degrees and correlation rate, we can find variable degrees and compression rate of main source that maximizes the rate.

The rest of paper is organized as follows: In Section II, we review the optimal rate problem for single user mode in BEC, BSC and AWGN channels. In Section III, the system model and its constraints are provided. Section IV is devoted to present the form of source-channel optimal rate LDPC code design problem for BE-MAC. A specific case of the system model in BEC is considered in Section V. Finally, the simulation results are given in Section VI. Conclusion is presented in Section VII.

## II. PRELIMINARIES: SINGLE USER OPTIMAL RATE CODE DESIGN

For an LDPC code with parity and variable degree distributions $\rho(x) = \sum_{j=2}^{D_c} \rho_j x^{j-1}$ and $\lambda(x) = \sum_{i=2}^{D_v} \lambda_i x^{i-1}$ respectively, the design rate of the code is:

$$R = 1 - \frac{\sum_{j=2}^{D_c} \rho_j/j}{\sum_{i=2}^{D_v} \lambda_i/i} \quad (1)$$

The DE constraint, must be forced to degree distributions, is changed from one channel to another one. For example, the DE constraint in BEC is:

$$\epsilon\lambda(1 - \rho(1-x)) \leq x \quad \forall x \in [0,\epsilon] \quad (2)$$

and, in BSC, it is:

$$(1-p)\lambda\left(\frac{1-\rho(1-2x)}{2}\right) + p\left[1 - \lambda\left(\frac{1+\rho(1-2x)}{2}\right)\right] \leq x \quad \forall x \in [0,p]. \quad (3)$$

By using [5], in AWGN channel it is:

$$x \geq \sum_i \lambda_i \phi\big(s + (i-1)\sum_j \rho_j \phi^{-1}(1-(1-r)^{j-1})\big). \quad (4)$$

(2) and (3) are polynomial with respect to x while the (4) is not.

The optimal rate problem for the BEC channel is achieved by solving the following nonlinear optimization problem:

$$\max \sum \frac{\lambda_i}{i} \quad (5)$$

Subject to: $\sum \lambda_i = 1$,

$\lambda_i \geq 0$,

$$\varepsilon\lambda(1 - \rho(1 - x)) \leq x \quad \forall x \in [0,1].$$

This problem is semi-infinite programming because the DE constraint includes infinite number of constraints. In [11-13], the authors showed for the DE constraint, the problem (5) can be reformulated as a Semi-definite Programming (SDP) problem. By this reformulation, one can find solution of the optimal rate problem.

Let $P(x) = x - \varepsilon\lambda(1 - \rho(1 - x))$, $q = \deg(P(x))$ and $\Pi(x) = (1 + x^2)^q P\left(\frac{x^2}{1+x^2}\right) = \sum_{j=0}^{2q} \Pi_j x^j$. Then, the SDP reformulation of the optimal rate problem in BEC is as follows [11-13]:

$$\max \sum \frac{\lambda_i}{i} \quad (6)$$

Subject to: $\sum \lambda_i = 1$,

$\Pi_l = \sum_{i+j=l} B_{ij} \quad \forall l = 0, \ldots, 2q$,

$B = (B_{ij}) \succcurlyeq 0, \ \lambda_i \geq 0.$

Note that the coefficients $\Pi_j$ are linear respect to $\lambda_i$s, and the operator $\succcurlyeq$ defines positive semi-definite matrices.

### III. SYSTEM MODEL AND SOURCE BSC CORRELATION

In this section, we focus on the problem of transmitting outputs of a correlated source over multiple access channel and binary alphabet, $U_1$, and compressed rates, $R_{s1}$, through related erasure channels with fixed capacity, $(C_1, C_2)$.

#### A. Region Slepian-Wolf

In [14], Slepian and Wolf showed the distributed source coding problem of correlated signals such as $(U_1, U_2)$, the achievable rate is given by:

$$R_{s1} \geq H(U_1 | U_2) \quad (7)$$
$$R_{s2} \geq H(U_2 | U_1) \quad (8)$$
$$R_{s1} + R_{s2} \geq H(U_1, U_2) \quad (9)$$

#### B. Region MAC

In the considered model, two users transmit correlated bits to a common receiver. In this case, the achievable rate region is given by [15]:

$$R_1 \leq I(X_1; Y | X_2) \quad (10)$$
$$R_2 \leq I(X_2; Y | X_1) \quad (11)$$
$$R_1 + R_2 \leq I(X_1, X_2; Y) \quad (12)$$

It is notable that we consider that $R_{s2} = 1$ which means no compression is used.

### IV. SOURCE AND CHANNEL OPTIMAL RATE CODE DESIGN

Now, we present our way for finding source-channel optimal rates for our proposed model. Considering (1), maximizing the channel code rate for single user can be described as follows [13]: Minimizing the linear function $\sum_{j=2}^{D_c} \rho_j/j$ subject to DE inequality and related constraints as follows:

$$\min \sum \frac{\rho_i}{i} \quad (16)$$

Subject to: $\sum \rho_i = 1$,

$\rho_i \geq 0$,

$\rho(1 - \varepsilon\lambda(x)) \geq 1 - x \quad \forall x \in [0,1].$

Now, consider two channels with erasure parameters $(\varepsilon_1, \varepsilon_2)$ and corresponding check degree distributions $\rho^{(1)}(t)$ and $\rho^{(2)}(t)$. For given check degree distributions, the aim in the optimal rate problem is to find variable degree distribution as well as their compression rates $R_{s1}$ in order to maximize the function $R_1 + R_2$, where $R_1$ and $R_2$ are given by (13) and (14).

In follow, we discuss about the decoder design of the source-channel LDPC code. Our decoder design for BE-MAC is similar to [16]. For decoding in BE-MAC, one is required a graph consisting of state nodes and an updating rule to present a way for joint decoding process. After receiving $(Y_1, Y_2)$ from channel and putting them into the joint decoder, the decoding is started. The process is done in two sub-decoders, each of them will be sum-product decoder or min-sum decoder. In addition, each sub-decoder transmits the side information bits and helps other sub-decoder. Different rules can be used for updating the side information which give us different DE inequalities. The result of this joint decoder is a code like Y which is derived from the pair $(Y_1, Y_2)$.

Therefore, the maximum rate problem for source channel optimal rate LDPC code design for one sender in BE-MAC with sources correlation in general cased can be described as follows, $R_{c2}$ is fixed:

$$\max(R_1 + R_2) = \max(R_{s1} R_{c1}) \quad (17)$$

Subject to: $\rho^{(1)}\left(1 - \varepsilon_1 \lambda^{(1)}(t)\right) \geq 1 - t \quad 0 \leq t \leq 1$,

$R_{s1} R_{c1} \leq I(X_1; Y | X_2)$,

$R_{s1} R_{c1} \leq I(X_1, X_2; Y) - R_{c2}$,

$H(U_1 | U_2) \leq R_{s1} \leq 1$,

$R_{s1} \geq H(U_1, U_2) - 1$,

$\sum \rho_i^{(1)} = 1, \ 0 \leq t \leq 1, \ \rho_i^{(1)} \geq 0$

In the next section, we focus on a specific case of the problem which source have been correlated by BSC model with other source. In this case, a semi-definite reformulation of the problem is provided too.

## V. Source and channel optimal rate code design: An specific case

In this section, an SDP reformulation of the problem (17) is given. This source correlation is a practical way for implementation Slepian-Wolf problem.

### A. BSC Correlation

Consider two binary source, $(U_1, U_2)$ and a binary random sequence Z where $U_2 = U_1 \oplus Z$. Let

$$\Pr(U_1 = U_2) = \Pr(Z = 0) = p. \qquad (18)$$

### B. The optimal rate problem for the correlated sources MAC

Let correlated information bits are sent over two channels after separate encoding process. Now, two received sequences $(Y_1, Y_2)$ in a joint decoder help us to recover Y.

## VI. Simulation results

In this section, we present some numerical results. In these simulations, regular variable check node degree distributions are considered.

**Example1**: For $\lambda^{(1)}(t) = t^{d_{v1}-1}$, $\lambda^{(2)}(t) = t^5$, $\rho^{(1)}(t) = 0.5821t^2 + 0.4179t^3$, $\rho^{(2)}(t) = 0.5821t^2 + 0.4179t^3$, $\varepsilon_1 = \varepsilon_2 = 0.3$ and $p = 0.89$, we have obtained the compression rates of Table 1 from solving the problem (32) by CVX package.

| $d_{v1} =$ | 5 | 6 | 7 | 8 | 9 | 10 | 11 |
|---|---|---|---|---|---|---|---|
| $R_{s1}$ | 1 | 0.5651 | 0.4999 | 0.4999 | 0.4999 | 0.4999 | 0.4999 |

TABLE I. Values of $R_{s1}$ for the problem set of Example 1

## VII. Conclusion

In this paper, we consider the joint source-channel model to design optimal codes which not only give optimal channel codes but also find the rate of compression for each link. Our focus is on the correlated source and multiple access channels. An SDP reformulation of the problem is proposed which can be solved in polynomial time by using interior point methods. Simulation results show that our works achieve the whole of capacity region for different manners of problem setup. Some extensions for this system model such as multiple accesses for more senders, other channels and so on can be extended in future works.

## References

[1] R. G. Gallager, "Low-Density Parity-Check Codes", Cambridge, MA:MIT Press, 1963.

[2] D. J. C. Mackay and R. M. Neal, "Near Shannon limit performance of low density parity check codes," IEE Elec. Lett., vol. 32, no. 18, pp. 1645–1646, Aug. 1996.

[3] T. Richardson and R. Urbanke, The capacity of low-density parity-check codes under message-passing decoding, IEEE Trans. on IT, vol. 47, no. 2, pp. 599–618, February 2001.

[4] T. Richardson, A. Shokrollahi and R. Urbanke, Design of capacity approaching irregular low-density parity-check codes, IEEE Trans. on IT, vol. 47, no. 2, pp. 619–637, February 2001.

[5] T. Richardson and R. Urbanke, Modern Coding Theory, Cambridge University Press, 2008. [Online]. Available: http://lthcwww.epfl.ch/mct/index.php.

[6] M. G. Luby, M. Mitzenmacher, M. A. Shokrollahi and D. A. Spielman, Efficient erasure-correcting codes, IEEE Trans. on IT, vol. 47, no. 2, pp. 569–584, February 2001.

[7] C. Di, D. Proietti, I. E. Telatar, T. J. Richardson, and R. L. Urbanke, Finite length analysis of low-density parity-check codes on the binary erasure channel, IEEE Trans. IT, vol. 48, no. 6, pp. 1570-1579, June 2002.

[8] T. J. Richrdson, A. Shokrollahi, and R. Urbanke, Finite-length analysis of various low-density parity-check ensembles for the binary erasure channel, in Proc. IEEE ISIT, Switzerland, July 2002.

[9] D. J. Costello and G. D. Forney, Channel coding: The road to channel capacity, Proceedings of the IEEE, vol. 95, no. 6, pp. 1150–1177, June 2007.

[10] Special issue on Codes on Graphs and Iterative Algorithms, IEEE Trans. on IT, vol. 47, no. 2, February 2001.

[11] H. Tavakoli, M. Ahmadian Attari, M.R. Peyghami, Optimal rate for irregular LDPC codes in binary erasure channel, ITW 2011, Paraty, Brasil, October 16-20, pp. 125-129, 2011.

[12] H. Tavakoli, M. Ahmadian Attari, M.R. Peyghami, Optimal rate and maximum erasure probability LDPC codes in binary erasure channel, 49th Allerton Conference, September 28-30, Illinois, 2011.

[13] H. Tavakoli, M. Ahmadian Attari, M.R. Peyghami, Optimal rate irregular low-density parity-check codes in binary erasure channel, IET Communications, Volume 6, issue13, 5 September 2012, p. 2000 – 2006

[14] D. Slepian, and J. K. Wolf, A coding theorem for multiple access channels with correlated sources, Bell System Tech. J., vol. 52, pp. 1037-1076, Sept. 1973.

[15] T.M. Cover and J.A. Thomas. Elements of Information Theory. Wiley Interscience, 1991.

[16] C. Kelley, D. Sridhara, LDPC Coding for the Three Terminal Erasure Relay Channel. Proc. of IEEE ISIT, July 9-14, 2006, Seattle, USA

[17] T. Cover, A. El Gamal, Capacity theorems for the relay channel, IEEE Trans IT, vol. 25, no. 5,1979,p. 572-584.